\documentclass[11pt,a4paper]{article}
\usepackage{amsmath}

\usepackage[psamsfonts]{amssymb}
\usepackage{amsmath}
\usepackage{epsfig}

\author{H. Mohseni Sadjadi\footnote{mohseni@phymail.ut.ac.ir; mohsenisad@ut.ac.ir}
\\ {\small Department of physics, University of Tehran ,}
\\ {\small P.O.B. 14395-547, Tehran 14399-55961, Iran}}
\title{ A Note on Gravitational Baryogenesis}

\begin{document}
\maketitle
\begin{abstract}
The coupling between Ricci scalar curvature and the baryon number
current dynamically breaks CPT in an expanding universe and leads
to baryon asymmetry. We study the effect of time dependence of
equation of state parameter of the FRW universe on this asymmetry.
\end{abstract}
\section{Introduction}
The origin of the difference between the number density of baryons
and anti-baryons is still an open problem in particle physics and
cosmology. The measurements of cosmic microwave background
\cite{CMB}, the absence of $\gamma$ ray emission from matter-
antimatter annihilation \cite{gamma} and the theory of Big Bang
nucleosynthesis \cite{Big} indicate that there is more matter than
antimatter in the universe. Observational results yield that the
ratio of the baryon number to entropy density is approximately
$n_b/s\sim 10^{-10}$. In \cite{sak}, it was pointed out that a
baryon-generating interaction must satisfy three conditions in
order to produce baryons and antibaryons at different rates: (1)
baryon number violation; (2) C and CP violation; (3) departure
from thermal equilibrium.

In \cite{coh}, by introducing a scalar field coupled to baryon
number current it was suggested that the baryon asymmetry may be
generated in thermal equilibrium while the CPT invariance is
dynamically violated. Similarly , in \cite{david}, by introducing
an interaction between Ricci scalar curvature and baryon number
current which dynamically violates CPT symmetry in expanding
Friedman Robertson Walker (FRW) universe, a mechanism for baryon
asymmetry was proposed. The proposed interaction shifts the energy
of a baryon relative to that of an antibaryon, giving rise to a
non-zero baryon number density in thermal equilibrium. The model
suggested in \cite{coh,david} was the subject of several studies
on gravitational baryogenesis and leptogenesis in different models
of cosmology \cite{theothers}.

But, in \cite{david}, the problem was restricted to the cases that
the equation of state parameter of the universe, $\omega$, is a
constant, and the role of time dependence of $\omega$ in
baryogenesis was neglected. As a consequence, in this scenario,
the baryon number asymmetry cannot be directly generated in
radiation dominated epoch. But in \cite{lamb}, in the framework of
modified theories of gravity, following the method of
\cite{david}, it was shown that the baryon asymmetry may be
generated even in the radiation dominated era.

In this paper, like \cite{david}, we assume that the universe is
filled with perfect fluids such as the scalar inflaton field and
radiation. The time dependence of $\omega$ is due to the fact
that: (1) these components have different equation of state
parameters, (2) they interact with each other and, (3) they may
have time dependent equation of state parameters.  We will study
the effect of each of these subjects on time derivative of the
Ricci scalar and consequently on baryogenesis. We will elucidate
our discussion through some examples.

Natural units $\hbar=c=k_B=1$ are used throughout the paper .

\section{The r\^ole of $\dot{\omega}$ in gravitational baryogenesis}

The mechanism proposed in \cite{david} to generate baryon
asymmetry is based on the introduction of a CP violating
interaction between the derivatives of the Ricci scalar and the
baryon number current, $J^\mu$, given by
\begin{equation}\label{1}
{\varepsilon\over M_{\star}^2}\int d^4x\sqrt{-g}(\partial_\mu R)
J^\mu,
\end{equation}
where $M_{\star}$ is a cutoff characterizing the scale of the
energy in the effective theory and $\varepsilon=\pm 1$. To obtain
the chemical potential for baryon ($\mu_B$) and antibaryons
($\mu_{\bar{B}}$) for spatially constant $R$, we use
\begin{equation}\label{2}
{1\over M_\star^2}(\partial_\mu R)J^\mu={1\over
M_\star^2}\dot{R}(n_B-n_{\bar B}),
\end{equation}
where $n_B$ and $n_{\bar{B}}$ are the baryon and antibaryon number
densities respectively. Therefore the energy of a baryon is
shifted by an amount of ${2\varepsilon\dot{R}\over M_\star^2}$
relative to an antibaryon. Thereby the interaction (\ref{1})
dynamically violates CPT. We can assign a chemical potential to
baryons: $\mu_B=-\mu_{\bar{B}}=-\varepsilon{\dot{R}\over
M_\star^2}$. So, in thermal equilibrium there will be a nonzero
baryon number density given by :
\begin{equation}\label{0}
n_b=n_B-n_{\bar{B}}={g_bT^3\over 6\pi^2}\left(\pi^2{\mu_B\over
T}+({\mu_B\over T})^3\right),
\end{equation}
where $g_b\sim \mathcal{O}(1)$ is the number of internal degrees
of freedom of baryons. The entropy density of the universe is
given by $s={2\pi^2\over 45} g_sT^3$, where $g_s\simeq 106$
indicates the total degrees of freedom for relativistic particles
contributing to the entropy of the universe \cite{kolb}. In the
expanding universe the baryon number violation decouples at a
temperature denoted by $T_D$ and a net baryon asymmetry remains.
The ratio ${n_b\over s}$ in the limit $T\gg m_b$ ($m_b$ indicates
the baryon mass), and $T\gg \mu_b$  is then:
\begin{equation}\label{3}
{n_b\over s}\simeq -\varepsilon{15g_b\over 4\pi^2g_s}{\dot{R}\over
M_\star^2T}|_{T_D}.
\end{equation}
Note that in different models we may have $\dot{R}<0$ as well as
$\dot{R}>0$, therefore introduction of $\varepsilon$ gives us the
possibility to choose the appropriate sign for $n_b$.

The geometry of the universe is described by the spatially flat
FRW metric
\begin{equation}\label{4}
ds^2=dt^2-a^2(t)(dx^2+dy^2+dz^2),
\end{equation}
where $a(t)$ is the scale factor. The Hubble parameter,
$H={\dot{a}\over a}$, satisfies
\begin{eqnarray}\label{5}
H^2&=&{8\pi G\over 3}\rho \nonumber \\
\dot{H}&=&-4\pi G(P+\rho).
\end{eqnarray}
$P$ and $\rho$ are the pressure and energy density. We assume that
the universe, filled with perfect fluids, satisfies the effective
equation of state: $P=\omega \rho$. The equation of state
parameter, $\omega$, can be expressed in terms of the Hubble
parameter:  $\omega=-1-{2\dot{H}\over 3H^2}$. The Ricci scalar
curvature is given by
\begin{eqnarray}\label{6}
R&=&-6H^2\left({\dot{H}\over H^2}+2\right)\nonumber \\
&=&-3H^2(1-3\omega).
\end{eqnarray}
From Eq. (\ref{5}), it follows that
\begin{equation}\label{7}
\dot{R}={\sqrt{3}\over M_p^3}(1+\omega)(1-3\omega)\rho^{3\over
2}+{3\over M_P^2}\rho \dot{\omega}.
\end{equation}
$M_p\simeq 1.22\times 10^{19}$ Gev is the Planck mass.  If
$\dot{\omega}=0$, Eq. (\ref{7}) reduces to the result obtained in
\cite{david}.

In the following we continue our study with a universe dominated
by two perfect fluids with equation of states $P_d=\gamma_d
\rho_d$ and $P_m=\gamma_m \rho_m$ interacting with each other,
through the source term $\Gamma_1\rho_d+\Gamma_2\rho_m$:
\begin{eqnarray}\label{8}
\dot{\rho_d}+3H(\rho_d+P_d)&=&\Gamma_1\rho_d+\Gamma_2\rho_m\nonumber
\\
\dot{\rho_m}+3H(\rho_m+P_m)&=&-\Gamma_1\rho_d-\Gamma_2\rho_m
\end{eqnarray}
Although these components don't satisfy the conservation equation
solely, but
\begin{equation}\label{9}
\dot{\rho}+3H(1+\omega)\rho=0.
\end{equation}
Note that $\rho\simeq \rho_m+\rho_d$ and $P\simeq P_m+P_d$.
$\omega$ can be written in terms of the ratio of energy densities,
$r={\rho_m\over \rho_d}$,
\begin{equation}\label{10}
\omega={\gamma_d+\gamma_m r\over 1+r}.
\end{equation}
From
\begin{equation}\label{11}
\dot{r}={\dot{\rho_m}\over \rho_d}-r{\dot{\rho_d}\over \rho_d},
\end{equation}
and Eq. (\ref{8}) one can determine the behavior of $r$ with
respect to the comoving time
\begin{equation}\label{12}
\dot{r}=-(r+1)(\Gamma_1+\Gamma_2r)+3Hr(\gamma_d-\gamma_m),
\end{equation}
or by suppressing $\gamma_d$
\begin{equation}\label{13}
\dot{r}=-(r+1)(\Gamma_1+\Gamma_2r)+3Hr(1+r)(\omega-\gamma_m).
\end{equation}
$\rho$ and $r$ may be related through
\begin{equation} \label{100}
{\dot{\rho}\over \rho}+{\dot{r}+(r+1)(\Gamma_1+\Gamma_2 r)\over
r(r+1)(\gamma_d-\gamma_m)}\left(\gamma_d+1+(\gamma_m+1)r\right)=0.
\end{equation}

Combining Eqs. (\ref{10}) and (\ref{11}) results in
 \begin{equation}\label{14}
 \dot{\omega}={\dot{\gamma_d}+r\dot{\gamma_m}\over 1+r}-{(\gamma_m-\gamma_d)
 (\Gamma_1+\Gamma_2r)\over
 1+r}-{3(\gamma_m-\gamma_d)^2Hr\over (1+r)^2}.
 \end{equation}
 The first and second terms show the effects of time dependence of
 $\gamma$'s, and interaction of fluid components on $\dot{\omega}$, respectively.
 Note that even for constant $\gamma$'s and in the absence of interactions,
 as the third term of (\ref{14}) indicates,  $\omega$ varies with time.
 This is due to the fact that the universe is assumed to be composed of
components with different equation of state parameters.  Putting
Eq. (\ref{14}) into Eq. (\ref{7}) gives
\begin{eqnarray}\label{15}
&&\dot{R}={\sqrt{3}\over
M_p^3}\Biggl({(1-3\gamma_m)(1+\gamma_m)r^2+2(1-\gamma_d-\gamma_m-3\gamma_d\gamma_m)r\over
(1+r)^2}+\nonumber \\
&& {(\gamma_d+1)(-3\gamma_d+1)\over (1+r)^2}\Biggr)\rho^{3\over
2}+{3\over M_p^2}\left({\dot{\gamma_d}+r\dot{\gamma_m}\over
 1+r}\right)\rho -\nonumber \\
 &&{3\over M_p^2}{(\Gamma_1+r\Gamma_2)(\gamma_m-\gamma_d)\over 1+r}
 \rho-{3\sqrt{3}\over M_p^3}{(\gamma_m-\gamma_d)^2r\over
 (1+r)^2}\rho^{3\over 2}.
 \end{eqnarray}
This equation can be rewritten in terms of $\omega$
\begin{eqnarray}\label{16}
&&\dot{R}={\sqrt{3}\over M_p^3}(1+\omega)(1-3\omega)\rho^{3\over
2}+{3\over
M_p^2}{(\gamma_m-\omega)\dot{\gamma_d}+(\omega-\gamma_d)\dot{\gamma_m}\over
\gamma_m-\gamma_d}\rho-\nonumber \\
&&{3\over
M_p^2}\left((\gamma_m-\omega)\Gamma_1+(\omega-\gamma_d)\Gamma_2\right)
\rho-{3\sqrt{3}\over
M_p^3}(\omega-\gamma_m)(\gamma_d-\omega)\rho^{3\over 2}.
\end{eqnarray}
If we neglect $\dot{\omega}$, as mentioned before, only the first
term remains: $\dot{R}=  {\sqrt{3}\over
M_p^3}(1+\omega)(1-3\omega)\rho^{3\over 2}$ which is zero at
$\omega={1\over 3}$ and at $\omega=-1$. But by taking
$\dot{\omega}$ into account we may have baryon asymmetry at
$\omega={1\over 3}$ and $\omega=-1$. Although the asymmetry
generated during inflation will be diluted away.

It is worth to study what happens when one of the fluid components
corresponds to radiation (e.g. produced after the inflation
epoch). To do so, we take $\lambda_m={1\over 3}$. In this case Eq.
(\ref{15}) reduces to
\begin{equation}\label{17}
\dot{R}={\sqrt{3}\over
M_p^3}{(1-3\gamma_d)(1+\gamma_d)\over(1+r)}\rho^{3\over 2}-{1\over
M_p^2}{(1-3\gamma_d)(\Gamma_1+\Gamma_2r)\over 1+r}\rho+{3\over
M_p^2}{\dot{\gamma_d}\over
 1+r}\rho.
\end{equation}
In general $\Gamma$'s may also be time dependent \cite{notari},
e.g., one can consider $\Gamma_1=\lambda_1H$ and
$\Gamma_2=\lambda_2H$ , where $\lambda_1,\lambda_2 \in \Re$
\cite{inter}. Depending on the model under consideration, the
third term of (\ref{17}), including the time derivative of
$\gamma_d$, may be simplified in terms of other parameters of the
model.  For example consider a massive scalar field of mass $m$,
with a time dependent equation of state parameter interacting with
radiation. The energy density, $\rho_d$, and the pressure, $P_d$,
of the scalar field satisfy
\begin{eqnarray}\label{18}
\rho_d&=&{1\over 2}\dot{\phi}^2+{1\over 2}m^2\phi^2 \nonumber \\
P_d&=&{1\over 2}\dot{\phi}^2-{1\over 2}m^2\phi^2.
\end{eqnarray}
The time dependent equation of state parameter of the scalar field
is then
\begin{equation}\label{19}
\gamma_d={{1\over 2}\dot{\phi}^2-{1\over 2}m^2\phi^2\over {1\over
2}\dot{\phi}^2+{1\over 2}m^2\phi^2}.
\end{equation}
The scalar field interacts with another component (radiation)
characterized by the equation of state parameter $\gamma_m={1\over
3}$:
\begin{eqnarray}\label{20}
\dot{\rho}_d+3H(\rho_d+P_d)&=&\Gamma_1\rho_d+\Gamma_2\rho_{\mathcal{R}} \nonumber \\
\dot{\rho_{\mathcal{R}}}+4H\rho_{\mathcal{R}}&=&-\Gamma_1\rho_d-\Gamma_2\rho_{\mathcal{R}}
\end{eqnarray}
The subscript $\mathcal{R}$ denotes the component with
$\gamma_m={1\over 3}$ . To determine $\dot{\gamma_d}$ in Eq.
(\ref{17}), let us define $u=(1-\gamma_d){\rho_d\over 2}$. Then
using
\begin{equation}\label{21}
\dot{u}=m\rho_d\sqrt{1-\gamma_d^2},
\end{equation}
which was verified in \cite{sad}, we are led to
\begin{equation}\label{22}
\dot{\gamma_d}=-2m\sqrt{1-\gamma_d^2}+(1-\gamma_d)(\Gamma_1+r\Gamma_2)-3H(1-\gamma_d^2).
\end{equation}
By supposing
\begin{equation}\label{23}
\Gamma_1=3\lambda_dH,\,\,\,
\Gamma_2=3\lambda_{\mathcal{R}}H,\,\,\,
\lambda_d,\lambda_{\mathcal{R}}\in \Re,
\end{equation}
we arrive at
\begin{equation}\label{24}
\dot{\gamma_d}=-2m\sqrt{1-\gamma_d^2}+3H(1-\gamma_d)(\lambda_d+\lambda_{\mathcal{R}}r-\gamma_d-1).
\end{equation}
Finally we deduce
\begin{equation}\label{25}
\dot{R}=-{6m\over M_p^2}{\sqrt{1-\gamma_d^2}\over
1+r}\rho+{2\sqrt{3}\over
M_p^3}{(-1-\gamma_d+\lambda_d+r\lambda_{\mathcal{R}})\over
1+r}\rho^{3\over 2}.
\end{equation}
If the potential is negligible with respect to the kinetic energy,
$\dot{\phi}^2\gg m^2\phi^2$, we have $\gamma_d\simeq 1$ and
$\dot{\gamma}_d\simeq 0$. In this case $\dot{R}={2\sqrt{3}\over
M_p^3}\left({-2+\lambda_d+r\lambda_\mathcal{R}\over
1+r}\right)\rho_d^{3\over 2}$.

When the scalar field dominates, i.e., in the limit $r\to 0$,
$\dot{R}=-{6m\over M_p^2}\sqrt{1-\gamma_d^2}\rho_d+{2\sqrt{3}\over
M_p^3}(-1-\gamma_d+\lambda_d)\rho_d^{3\over 2}$.

In the limit $r\to \infty$, i.e, when $\rho_{\mathcal{R}}$
dominates
\begin{equation}\label{102}
\dot{R}={2\sqrt{3}\over M_p^3}
\lambda_{\mathcal{R}}\rho_{\mathcal{R}}^{3\over 2}.
\end{equation}
So, even in the radiation dominated era, we can have $\dot{R}\neq
0$ if $\lambda_\mathcal{R}\neq 0$. For radiation component, the
energy density is related to the equilibrium temperature via
\cite{kolb}
\begin{equation}\label{32}
\rho_{\mathcal{R}}=\epsilon_{\mathcal{R}}T^4,
\end{equation}
where $\epsilon_{\mathcal{R}}={\pi^2\over 30}g_{\star}$ and
$g_\star$ counts the total number of effectively massless degrees
of freedom. Its magnitude is  $g_\star\simeq g_s$. Therefore the
baryon asymmetry in terms of temperature can be determined from
(\ref{3}), (\ref{102}), and (\ref{32}):
\begin{equation}\label{103}
{n_b\over s}\simeq 2.5  g_b \lambda_{\mathcal{R}}{T_D^5\over
M_\star^2M_p^3}.
\end{equation}
By defining $M_\star=\alpha M_p$ we find that ${n_b\over s}$ is of
order
\begin{equation} {n_b\over s}\sim {\lambda_\mathcal{R}\over
\alpha^2}\left({T_D\over M_p}\right)^5.
\end{equation}

When $\gamma$'s are time dependent, in general, derivation of the
exact solutions of Eq. (\ref{20}) is not possible. Thus explaining
the precise behavior of the fluid components in terms of
temperature, except in special situations like (\ref{103}),  is
not straightforward. But when $\gamma$'s are constant, it may be
possible to determine baryogenesis in terms of decoupling
temperature. We will study this situation through two examples.

As a first example consider $\gamma_m=\gamma_\mathcal{R}=1/3$. We
assume that $\gamma_d$ is a constant satisfying $\gamma_d>{1\over
3}$ corresponding to a non-thermal component decreasing more
rapidly than radiation \cite{david}. If there is no energy
exchange, i.e, $\Gamma_1=\Gamma_2=0$:
\begin{eqnarray}\label{26}
\dot{\rho_d}+3(\gamma_d+1)H\rho_d&=&0
\nonumber \\
\dot{\rho_\mathcal{R}}+4H\rho_\mathcal{R}&=&0.
\end{eqnarray}
In this case
\begin{equation}\label{27}
\dot{R}={\sqrt{3}\over
M_p^3}{(1-3\gamma_d)(1+\gamma_d)\over(1+r)}\rho^{3\over 2}.
\end{equation}
This result may reproduced by taking $\gamma_d=1$ (which results
in $\dot{\gamma_d}=0$) and $\lambda_{\mathcal{R}}=\lambda_d=0$ in
(\ref{25}).  $r$ satisfies
\begin{equation}\label{28}
\dot{r}=Hr(3\gamma_d-1),
\end{equation}
hence, $r$ is an increasing function of comoving time, i.e.,
$\rho_d$ decreases faster than the radiation component. $\dot{R}$
is also an increasing function of time. To see this let us
consider
\begin{equation}\label{31}
\ddot{R}={1\over
2M_p^4}(1+\gamma_d)(3\gamma_d-1)\left(9(\gamma_d+1)+2r(3\gamma_d+5)\right){\rho^2\over
(1+r)^2}.
\end{equation}
which was obtained using Eqs. (\ref{100}) and (\ref{27}). Hence
$\gamma_d>{1\over 3}$ leads to $\ddot{R}>0$.

Eq. (\ref{26}) implies $\rho_{\mathcal{R}}\propto a^{-4}$, hence
the temperature redshifts as $T\propto a^{-1}$, this relation and
Eq. (\ref{26}) imply $\rho_d\propto T^{3(1+\gamma_d)}$.  If at a
temperature denoted by $T=T_{RD}$, $\rho_\mathcal{R}=\rho_d$, we
can write
\begin{equation}\label{33}
\rho_d=\epsilon_{\mathcal{R}} T_{RD}^4\left({T\over
T_{RD}}\right)^{3(1+\gamma_d)},
\end{equation}
which results in
\begin{equation}\label{34}
\rho=\epsilon_{\mathcal{R}} T_{RD}^4\left(\left({T\over
T_{RD}}\right)^4+\left({T\over
T_{RD}}\right)^{3(1+\gamma_d)}\right),
\end{equation}
and
\begin{equation}\label{35}
r=\left({T\over T_{RD}}\right)^{1-3\gamma_d}.
\end{equation}
by putting Eqs. (\ref{34}) and (\ref{35}), in Eq. (\ref{27}) and
subsequently in Eq. (\ref{3}) at $T=T_D$, the baryon asymmetry may
be determined:
\begin{equation}\label{36}
{n_b\over s}\simeq {1.28 g_b\over
M_{\star}^2M_p^3}(1+\gamma_d)(3\gamma_d-1){T_{RD}^6\over
T_D}\left({T_D\over T_{RD}}\right)^{9(1+\gamma_d)\over
2}\sqrt{1+\left({T_D\over T_{RD}}\right)^{1-3\gamma_d}}.
\end{equation}

In the limit $r\to 0$ (related to $T_D\gg T_{RD}$),
\begin{equation}\label{37}
{n_b\over s}\simeq  {1.28 g_b\over
M_\star^2M_p^3}(3\gamma_d-1)(1+\gamma_d){T_{RD}^6\over
T_D}\left({T_D\over T_{RD}}\right)^{9(1+\gamma_d)\over 2},
\end{equation}
which is the same as the result obtained in \cite{david}, where
the effect of time dependence of $\omega$ was ignored, indeed in
this limit $\dot{\omega}$ does not contribute in baryon asymmetry.
In the limit $r\gg 1$ which corresponds to $\rho_d\ll
\rho_{\mathcal{R}}$ we have
\begin{equation}\label{38}
{n_b\over s}\simeq {1.28  g_b\over
M_\star^2M_p^3}(1+\gamma_d)(3\gamma_d-1){T_{RD}^6\over
T_D}\left({T_D\over T_{RD}}\right)^{5+3\gamma_d}.
\end{equation}
E. g. if we define $T_D=\beta T_{RD}$, $\beta\ll 1$, and as before
$M_\star=\alpha M_p$, then the order of magnitude of  ${n_b\over
s}$ is obtained as
\begin{equation} {n_b\over s}\sim
(3\gamma_d-1){\beta^{4+3\gamma_d}\over \alpha^2}\left(T_{RD}\over
M_p\right)^5.
\end{equation}

As a second example consider a model,  with constant $\gamma_d$,
and $\gamma_m={1\over 3}$.  Also assume that at least one of the
constant $\Gamma$'s are not zero. Hence
\begin{equation}\label{39}
\dot{R}={\sqrt{3}\over
M_p^3}{(1-3\gamma_d)(1+\gamma_d)\over(1+r)}\rho^{3\over 2}-{1\over
M_p^2}{(1-3\gamma_d)(\Gamma_1+\Gamma_2r)\over 1+r}\rho,
\end{equation}
may be used to determine ${n_b\over s}$. In this situation
although Eq. (\ref{32}) holds but Eq. (\ref{33}) is no longer
valid. The temperature does not  behave as $T\propto a^{-1}$ and,
instead, we have $T=f(a)$ \cite{kolb}, \cite{dav}.  In some
special cases, $\dot{R}$ can be easily determined in terms of
decoupling temperature. For example, at $r\simeq 1$ (when
$\omega\simeq {3\gamma_d+1\over 6}$) we have
\begin{equation}\label{40}
\rho\simeq 2\epsilon_{\mathcal{R}} T^4
\end{equation}
and
\begin{equation}\label{41}
{n_b\over s}\simeq {\varepsilon g_b\over M_\star^2}\left({3.6\over
M_p^3}(3\gamma_d-1)(1+\gamma_d)T^5_D-{0.25\over
M_p^2}(3\gamma_d-1)(\Gamma_1+\Gamma_2)T_D^3\right).
\end{equation}
Note that in the above example $T_D=T_{RD}$.

Also For $r\to 0$ and $r\to \infty$, corresponding to
$\omega\simeq \gamma_d$ and $\omega\simeq \gamma_m$ respectively,
Eq. (\ref{14}) leads to
\begin{equation}\label{42}
\dot{\omega} = \left\{ \begin{array}{ll}
(\gamma_d-\gamma_m)\Gamma_1, \quad r\to 0\\
(\gamma_d-\gamma_m)\Gamma_2,  \quad  r\to \infty\\
\end{array} \right.
\end{equation}
Therefore for $\gamma_m={1\over 3}$,
\begin{equation}\label{43}
\dot{R} = \left\{ \begin{array}{ll}
{\sqrt{3}\over {M_p^3}}(1+\gamma_d)(1-3\gamma_d)\rho_d^{3\over 2}
-{(1-3\gamma_d)\Gamma_1\over M_p^2}\rho_d,\quad r\to 0\\
{(3\gamma_d-1)\Gamma_2\over M_p^2}\rho_{\mathcal{R}},\quad  r\to \infty.\\
\end{array} \right.
\end{equation}
E.g., if one suppose that the reheating process after the
inflation is due to the decay of a  scalar field (the inflaton
field: $\phi$) to (ultra)relativistic species, via coherent
oscillation \cite{kolb}, then
\begin{eqnarray}\label{44}
\dot{\rho}_{\phi}+3H\rho_{\phi}^2&=&-\Gamma_\phi
\rho_\phi\nonumber
\\
\dot{\rho}_R+4H\rho_{\mathcal{R}}&=&\Gamma_\phi\rho_\phi,
\end{eqnarray}
where $\rho_{\mathcal{R}}$ is the energy density in the
relativistic decay products. If $M^4$ is the vacuum energy of the
scalar field at the beginning of the oscillation, from $t\simeq
{M_p\over M^2}$ until $t\simeq \Gamma_\phi^{-1}$, $\phi$
particles, which behave like non-relativistic particles (i.e.,
$\gamma_\phi=0$), dominate the energy density \cite{kolb}. During
this time, $a(t)\propto t^{2\over 3}$, $\rho\propto a^{-3}$ and
\begin{equation}\label{45}
\dot{R}={\sqrt{3}\over M_p^3}\rho_\phi^{3\over
2}+{\Gamma_\phi\over M_p^2}\rho_\phi.
\end{equation}
Although $\rho_{\mathcal{R}}\propto T^4$ and Eq. (\ref{32}) is
still valid, but due to the interaction term,
$\rho_{\mathcal{R}}\propto a^{-4}$ does not hold. Then $T\propto
a^{-1}$ is not valid. In this case one can show that the
temperature can be approximated by \cite{dav}
\begin{equation}\label{46}
T=T_{RD}\left({a_{RD}\over a}\right)^{3\over 8}.
\end{equation}
Hence at the time of decoupling
\begin{equation}\label{47}
\rho_\phi=\epsilon_{\mathcal{R}} T_{RD}^4\left({T_D\over
T_{RD}}\right)^8,
\end{equation}
and
\begin{equation}\label{48}
{n_b\over s}\simeq {\varepsilon\over M_\star^2}{g_b\over 8}\left({
\sqrt{3\epsilon_\mathcal{R}}\over
M_p^3}T_{RD}^{-1}T_D^{6}+{\Gamma_\phi\over
M_p^2}T_{RD}T_D^2\right).
\end{equation}
To derive the above equation we have multiplied our result by the
dilution factor $\left({T_{RD}\over T_D}\right)^5$ \cite{david}.
The first term is the same as the result obtained in \cite{david},
where $\dot{\omega}$ was ignored. If we take $\Gamma_\phi\sim
{T_{RD}^2\over M_p}$, and as before define $M_\star=\alpha M_p$
and $T_D=\beta T_{RD}$, then we are able to compare the order of
magnitudes of terms contributing in (\ref{48}):
\begin{equation}
{1\over M_\star^2}{g_b\over 8}\left({
\sqrt{3\epsilon_\mathcal{R}}\over
M_p^3}T_{RD}^{-1}T_D^{6}\right)\sim 10{\beta^6\over
\alpha^2}\left(T_{RD}\over M_p\right)^5,
\end{equation}
\begin{equation}
 {1\over
M_\star^2}{g_b\over 8}\left({\Gamma_\phi\over
M_p^2}T_{RD}T_D^2\right)\sim {\beta^2\over
\alpha^2}\left({T_{RD}\over M_p}\right)^5.
\end{equation}
When $\beta \gg 1$ the second term in (\ref{48}) which comprises
the effect of decay width is negligible.

At the end let us note that if $\omega$ were time independent,
${n_b\over s}$ would be vanish in radiation dominated era. But by
taking $\dot{\omega}$ into account, it is easy to see that if
$\Gamma_2\neq 0$, in the limit $\rho_{\mathcal{R}}\gg \rho_\phi$
we have $\dot {R}={(3\gamma_d-1)\Gamma_2\over
M_p^2}\rho_{\mathcal{R}}$, therefore
\begin{equation}\label{49}
{n_b\over s}\simeq -\varepsilon{g_b\over 8
M_\star^2}{(3\gamma_d-1)\Gamma_2\over M_p^2}T_D^3.
\end{equation}
Thereby ${n_b\over s}\sim {(3\gamma_d-1)\over \alpha^2}{\Gamma_2
T_D^3\over M_p^4}$.

\end{document}